\documentclass[aps,amsmath,preprint,epsf,superscriptaddress,nofootinbib]{revtex4-1}
\usepackage{graphicx}
\usepackage{bm}
\usepackage{color}
\usepackage{amsmath,amssymb}	
\usepackage{hyperref}
\usepackage{setspace}
\usepackage{longtable}
\usepackage{booktabs}
\usepackage{comment}
\usepackage{array}

\def\slashchar#1{\setbox0=\hbox{$#1$} 
\dimen0=\wd0 
\setbox1=\hbox{/} \dimen1=\wd1 
\ifdim\dimen0>\dimen1 
\rlap{\hbox to \dimen0{\hfil/\hfil}} 
#1 
\else 
\rlap{\hbox to \dimen1{\hfil$#1$\hfil}} 
/ 
\fi}

\def\a{\alpha}

\def\e{\epsilon}

\def\k{\kappa}
\def\l{\lambda}

\def\L{\Lambda}

\def\beq{\begin{eqnarray}}
\def\eeq{\end{eqnarray}}

\newcommand{\gsim}{ \mathop{}_{\textstyle \sim}^{\textstyle >} }

\newcommand{\vev}[1]{ \left\langle {#1} \right\rangle }

\begin{document}
\newcolumntype{Y}{>{\centering\arraybackslash}p{20pt}} 


\preprint{IPMU18-0039}
\bigskip

\title{Gauged Peccei-Quinn Symmetry \\{\sl -- A Case of Simultaneous Breaking of SUSY and PQ Symmetry --}}

\author{Hajime Fukuda}
\email[e-mail: ]{hajime.fukuda@ipmu.jp}
\affiliation{Kavli IPMU (WPI), UTIAS, The University of Tokyo, Kashiwa, Chiba 277-8583, Japan}
\author{Masahiro Ibe}
\email[e-mail: ]{ibe@icrr.u-tokyo.ac.jp}
\affiliation{Kavli IPMU (WPI), UTIAS, The University of Tokyo, Kashiwa, Chiba 277-8583, Japan}
\affiliation{ICRR, The University of Tokyo, Kashiwa, Chiba 277-8582, Japan}
\author{Motoo Suzuki}
\email[e-mail: ]{m0t@icrr.u-tokyo.ac.jp}
\affiliation{Kavli IPMU (WPI), UTIAS, The University of Tokyo, Kashiwa, Chiba 277-8583, Japan}
\affiliation{ICRR, The University of Tokyo, Kashiwa, Chiba 277-8582, Japan}
\author{Tsutomu T. Yanagida}
\email[Hamamatsu Professor. e-mail: ]{tsutomu.tyanagida@ipmu.jp}
\affiliation{Kavli IPMU (WPI), UTIAS, The University of Tokyo, Kashiwa, Chiba 277-8583, Japan}

\date{\today}

\begin{abstract}
Recently, a simple prescription to embed the global Peccei-Quinn (PQ) symmetry into a gauged $U(1)$ symmetry has been proposed.
There, explicit breaking of the global PQ symmetry expected in quantum gravity are highly suppressed due to the gauged PQ symmetry.
In this paper, we apply the gauged PQ mechanism to models where the global PQ symmetry and supersymmetry (SUSY) are simultaneously broken 
at around $\mathcal{O}(10^{11-12})$\,GeV.
Such scenario is motivated by an intriguing coincidence between the supersymmetry breaking scale which explains the observed Higgs boson mass
by the gravity mediated sfermion masses, and the PQ breaking scale which evades all the astrophysical and the cosmological constraints.
As a concrete example, we construct a model which consists of a simultaneous supersymmetry/PQ symmetry breaking sector based on $SU(2)$ dynamics
and an additional PQ symmetry breaking sector based on $SU(N)$ dynamics.
We also show that new vector-like particles are predicted in the TeV range in the minimum model, which can be tested by the LHC experiments.
\end{abstract}

\maketitle

\section{Introduction}
\label{sec:intro}
The Peccei-Quinn (PQ) mechanism\,\cite{Peccei:1977hh,Peccei:1977ur,Weinberg:1977ma,Wilczek:1977pj} 
provides us with a very successful solution to the strong $CP$ problem.
The effective $\theta$-angle of QCD is canceled  by the vacuum expectation value (VEV) 
of the pseudo-Nambu-Goldstone boson, the axion $a$, which results from spontaneous breaking of 
the global $U(1)$ Peccei-Quinn symmetry, $U(1)_{PQ}$.

The solution of the strong $CP$ problem based on a global symmetry is, however, not on the very firm theoretical ground. 
As the QCD anomaly explicitly breaks the $U(1)_{PQ}$ symmetry, it cannot be an exact symmetry by definition. 
Besides, it is also argued that all global symmetries are broken by quantum gravity 
effects~\cite{Hawking:1987mz, Lavrelashvili:1987jg, Giddings:1988cx, Coleman:1988tj, Gilbert:1989nq, Banks:2010zn}.
The explicit breaking of the PQ symmetry easily spoils the success of the PQ mechanism.

In Ref.\,\cite{Fukuda:2017ylt}, a simple prescription has been proposed, with which
the global $U(1)_{PQ}$ symmetry is embedded into a ``gauged" $U(1)$ symmetry, $U(1)_{gPQ}$.
There, the anomalies of the gauged PQ symmetry are canceled between the contributions from two (or more) PQ charged sectors.
With appropriate charge assignment of $U(1)_{gPQ}$, the PQ charged sectors are highly decoupled with each other, 
and a global $U(1)_{PQ}$ symmetry appears as an accidental symmetry. 
As a part of the gauge symmetry, the accidental $U(1)_{PQ}$ is also well protected from  explicit breaking caused by quantum gravity effects.
This prescription  provides a concise generalization of previous attempts to achieve the PQ symmetry 
as an accidental symmetry resulting from (discrete) gauge symmetries~\cite{Barr:1992qq,Kamionkowski:1992mf,Holman:1992us,Dine:1992vx,Dias:2002gg,Carpenter:2009zs,Harigaya:2013vja,Harigaya:2015soa,Redi:2016esr,Duerr:2017amf}.

In this paper, we apply the construction of the gauged PQ symmetry to a model in which 
the global PQ symmetry and supersymmetry are simultaneously broken at around $\mathcal{O}(10^{11-12})$\,GeV~\cite{Feldstein:2012bu}.
Such scenario is motivated by an intriguing coincidence between the supersymmetry breaking scale which explains the observed Higgs boson mass
by the gravity mediated sfermion masses in the hundreds to thousands TeV range~\cite{Okada:1990vk} 
and the PQ breaking scale which evades all the astrophysical and the cosmological constraints.%
\footnote{For correspondence between the sfermion mass scale and the Higgs boson mass, 
see also~\cite{Ellis:1990nz,Okada:1990gg,Haber:1990aw}.
For constraints on the PQ breaking scale, see, e.g.,~\cite{Raffelt:2006cw,Kawasaki:2013ae,Patrignani:2016xqp}.}

The organization of the paper is as follows. 
In section~\ref{sec:prescription}, we summarize the supersymmetric version of the gauged PQ mechanism.
In section~\ref{sec:proto}, we construct a model in which supersymmetry and the PQ symmetry are 
broken simultaneously by $SU(2) $  strong dynamics.
In section~\ref{sec:model}, we apply the gauged PQ mechanism to the model of simultaneous symmetry breaking. 
The final section is devoted to our conclusions.


\section{General Prescription of the Gauged PQ Mechanism}
\label{sec:prescription}
In this section, we briefly summarize a supersymmetric version of the gauged PQ mechanism~\cite{Fukuda:2017ylt}. 
\subsection{Would-be Goldstone and Axion Superfields}
As a simple example, let us consider two global PQ symmetries $U(1)_{PQ_1}$ and $U(1)_{PQ_2}$, 
which are broken by the VEVs of $\Phi_1$, $\bar{\Phi}_1$ and 
$\Phi_2$, $\bar{\Phi}_2$, respectively.
For instance, such vacuum is achieved by the superpotential,
\beq
\label{eq:simple super}
W=\l_1 X_1(2\Phi_1 \bar{\Phi}_1-\Lambda_1^2)+\l_2 X_2(2\Phi_2 \bar{\Phi}_2-\Lambda_2^2)\ .
\eeq
Here, $\Phi_i$ and $\bar{\Phi}_i$ $(i=1,2)$ have charges $\pm 1$ under $U(1)_{PQ_i}$
and have vanishing charges under $U(1)_{PQ_j}$ $(j\neq i)$, respectively.
The superfields, $X_{1,2}$, have vanishing charges under both the PQ symmetries.
The parameters $\l_{1,2}$ are coupling constants, and $\L_{1,2}$ are dimensionful parameters.
After  the spontaneous breaking of the PQ symmetries, $\Phi$'s lead to the Goldstone superfields  $A_{1,2}$,%
\footnote{Here, we set the origins of $A_{i}$ at which $\Phi_{i} = \bar{\Phi}_i$, while $\vev{\Phi_i}\neq \vev{\bar{\Phi}_i}$ for $\vev{A_{i}}\neq 0$. }
\begin{eqnarray}
\label{eq:dec1}
\Phi_1 &=& \frac{1}{\sqrt 2}\L_1 e^{A_1/\L_1 }\ ,  \quad \bar\Phi_1 =  \frac{1}{\sqrt 2}\L_1 e^{-A_1/\L_1 }\ , \\
\label{eq:dec2}
\Phi_2 &=&  \frac{1}{\sqrt 2}\L_2 e^{A_2/\L_2 }\ ,  \quad \bar\Phi_2 =  \frac{1}{\sqrt 2}\L_2 e^{-A_2/\L_2 }\ .
\end{eqnarray}
By using the Goldstone superfields, the PQ symmetries are realized by,
\begin{eqnarray}
A_1/\L_1 & \to&   A_1/\L_1  + i \alpha_1 \ , \quad(\alpha_1 = 0-2\pi)\ , \\
A_2/\L_2  &\to&   A_2/\L_2  + i \alpha_2 \ , \quad(\alpha_2 = 0-2\pi)\ .
\end{eqnarray}
 
The PQ symmetries are communicated to the supersymmetric Standard Model (SSM) sector 
by introducing extra quark multiplets as in the KSVZ axion model~\cite{Kim:1979if, Shifman:1979if}.
Throughout this paper, we assume that the extra multiplets form ${\bf 5}$ and $\bar{\bf 5}$ representations 
of the $SU(5)$ gauge group of the  Grand Unified Theory (GUT).
Let us suppose that $\Phi_{1,2}$ couple to $N_1$ and $N_2$ flavors of the KSVZ extra multiplets 
${\bf 5}_i$, $\bar{\bf 5}_i$ $(i=1, 2)$, respectively,
\beq
W=\Phi_1 {\bf 5}_1\bar{\bf 5}_1 + \bar{\Phi}_2 {\bf 5}_2\bar{\bf 5}_2\ .
\eeq
Through the above coupling, both the global PQ symmetries are broken by the Standard Model anomaly.
The anomalous breaking of the global PQ symmetries lead to the anomalous coupling of the Goldstone superfields,
\begin{eqnarray}
\label{eq:anomSF}
W_{\rm anom} = \frac{1}{8\pi^2}\left(N_1\frac{A_1}{\Lambda_1} 
-N_2 \frac{A_2}{\Lambda_2}\right) \sum_l W_l^{\a}W_{l\a}\ ,
\end{eqnarray}
where, $W_l^{\a}$ $(l=1,2,3)$ denote the field strength superfields of the Standard Model gauge interactions.%
\footnote{Here, the gauge indices of $SU(3)_c$ and $SU(2)_L$ are suppressed, and the GUT normalization is used for $U(1)_Y$.}
We normalize the gauge field strength so that the gauge kinetic functions are given by
\beq
\mathcal{L} = \frac{1}{2i}\left[\sum \tau_l W_l^{\alpha} W_{l\alpha}\right]_F + h.c. \ ,
\eeq
with
\beq
\tau_l = \frac{i}{g_l^2} + \frac{\theta_l}{8\pi^2}\ ,
\eeq
where $g_l$ and $\theta_l$  are the gauge coupling constants and the vacuum angles of the corresponding gauge interactions.

An important observation here is that there is a linear combination of the PQ symmetries for which the Standard Model anomalies are absent.
In fact, a $U(1)$ symmetry under which  $\Phi_{1,2}$ have charges $q_1$ and $q_2$
is free from the Standard Model anomaly for
\begin{eqnarray}
\label{eq:anomfree}
q_1N_1-q_2N_2=0\ .
\end{eqnarray}
In the gauged PQ mechanism, we identify the anomaly-free combination to  be a gauge symmetry $U(1)_{gPQ}$.
The gravitational anomaly and the self-anomaly of the $U(1)_{gPQ}$ are canceled 
by adding $U(1)_{gPQ}$ charged singlet fields.
Hereafter, we take $q_1$ and $q_2$ are both positive and relatively prime numbers without loss of generality.

In the gauged PQ mechanism, one of the linear combinations of $A_{1,2}$ is the would-be Goldstone supermultiplet,
and the other combination corresponds to the physical axion superfield.
To see how the physical axion is extracted, let us consider the K\"ahler potential of  $\Phi$'s,
\beq
K=\Phi_1^{\dagger} e^{-2q_1gV}\Phi_1+\bar{\Phi}_1^{\dagger} e^{2q_1gV}\bar{\Phi}_1
+\Phi_2^{\dagger} e^{-2q_2 gV}\Phi_2+\bar{\Phi}_2^{\dagger} e^{2 q_2g V}\bar{\Phi}_2 \ ,
\eeq
where $V$ and $g$ are the $U(1)_{gPQ}$ gauge supermultiplet and the gauge coupling constant, respectively.
Under the $U(1)_{gPQ}$ gauge transformation, the gauge field is shifted by,
\begin{eqnarray}
2gV \to 2gV' = 2gV - i \Theta + i \Theta^\dagger \ ,
\end{eqnarray}
with $\Theta$ being the gauge parameter superfield.
By substituting Eqs.(\ref{eq:dec1}) and (\ref{eq:dec2}), the K\"ahler potential is reduced to
\beq
K&=& 
 \Lambda_1^2 \cosh\left({2q_1gV - \frac{A_1^\dagger + A_1}{\L_1} }\right) 
+ \Lambda_2^2 \cosh\left({2q_2gV - \frac{A_2^\dagger + A_2}{\L_2} }\right)\ .
\eeq

The physical axion and the would-be Goldstone superfields $A$ and $G$ are obtained by rearranging $A_{1,2}$ by
\begin{eqnarray}
\left(
\begin{array}{cc}
A^{(\dagger)}
   \\
G^{(\dagger)}
\end{array}
\label{eq:decomp}
\right)=
\frac{1}{\sqrt{q_1^2 \Lambda_1^2 + q_2^2\Lambda_2^2 }}\left(
\begin{array}{cc}
q_2 \Lambda_2   &  -q_1 \Lambda_1   \\
q_1 \Lambda_1  &   q_2 \Lambda_2
\end{array}
\right)
\left(
\begin{array}{cc}
A_1^{(\dagger)}   \\
A_2^{(\dagger)}
\end{array}
\right)\ .
\end{eqnarray}
By using $A$ and $G$, the K\"ahler potential is rewritten by,
\begin{eqnarray}
\label{eq:AGK}
K &=&
 \L_1^2 \cosh\left(
2q_1 \tilde V - \frac{2q_2}{m_V}\frac{\L_2}{\L_1} (A^{\dagger}+A)
\right)
+ 
\L_2^2 \cosh\left(
2q_2 \tilde V + \frac{2q_1}{m_V}\frac{\L_1}{\L_2} (A^{\dagger}+A)
\right) \ , 
\end{eqnarray}
where 
\begin{eqnarray}
\tilde V &=& V - \frac{g}{m_V}(G^\dagger+G) \ , \\
m_V &=& 2g\sqrt{q_1^2 \L_1^2 + q_2 \L_2^2}\ .
\end{eqnarray}
The final expression of Eq.\,(\ref{eq:AGK}) shows there is no bi-linear term which mixes $A$ and $\tilde{V}$.
Therefore, we find that $A$ corresponds to the physical axion superfield, while $G$ is 
the would-be Goldstone superfield  which is absorbed by ${V}$ in the unitarity gauge.
It should be noted that the physical axion $A$ is invariant under the gauge $U(1)_{gPQ}$ transformation.

For a later purpose, let us discuss the domain and the effective decay constant of the axion.
The domains of the imaginary parts of $A_{1,2}$ (corresponding to the phases of $\Phi_{1,2}$) 
are given by
\begin{eqnarray}
\frac{\operatorname{Im}[A_{i}]}{\L_{i}} = \frac{a_{i}}{f_{i}} = [0, 2\pi)\ , \quad (i = 1,2)\ ,
\end{eqnarray}
where $a_{i} = \sqrt{2}\operatorname{Im}[A_i]$ and $f_{i} = \sqrt{2}\Lambda_i$.
When $q_{1}$ and $q_2$ are relatively prime integers, the gauge invariant axion interval is given by~\cite{Fukuda:2017ylt},
\beq
a = \sqrt{2} \operatorname{Im}[A] = \left[0,~\frac{2\pi f_1 f_2}{\sqrt{q_1^2f_1^2+q_2^2f_2^2}} \right).
\eeq
Accordingly, the global $U(1)_{PQ}$ symmetry is realized by
\begin{eqnarray}
\frac{a}{F_a}  \to \frac{a'}{F_a}   = \frac{a}{F_a}  + \delta_{PQ}\ , \quad (\delta_{PQ} = 0-2\pi)\ ,
\end{eqnarray}
where $F_a$ is defined as an effective decay constant,
\beq
\label{eq:effF}
F_a=\frac{f_1f_2}{\sqrt{q_1^2f_1^2+q_2^2f_2^2}}
=\frac{\sqrt 2\Lambda_1\Lambda_2}{\sqrt{q_1^2\Lambda_1^2+q_2^2\Lambda_2^2}} \ .
\eeq

\subsection{Accidental Global PQ Symmetry}
As argued in~\cite{Hawking:1987mz, Lavrelashvili:1987jg, Giddings:1988cx, Coleman:1988tj, Gilbert:1989nq, Banks:2010zn},
global symmetries are expected to be broken by quantum gravity  effects which are manifested by
explicit breaking terms suppressed by the Planck scale.
On the other hand, such explicit breaking does not appear for the $U(1)_{gPQ}$ as it is an exact gauge symmetry.
Thus, question is  how well the accidental global PQ symmetry (corresponding to the shift of $A$) 
is protected from  explicit breaking by the $U(1)_{gPQ}$ symmetry.
An important observation here is that the $U(1)_{gPQ}$ symmetry is not distinguishable from the global PQ symmetries in each sector.
Thus, there are no explicit breaking terms of the global PQ symmetries which consist of the fields in each PQ symmetric sector. 
Therefore, the interaction terms which potentially ruin the global PQ symmetries are the gauge invariant operators consisting of 
the fields in multiple PQ symmetric sectors.
In the following, we estimate how badly the accidental global 	PQ symmetry is broken.

In the present model, the lowest dimensional $U(1)_{gPQ}$ invariant operators which break the global PQ symmetries are given by,
\beq
\label{eq:explicitW}
W\sim \frac{1}{M_{\rm PL}^{q_1+q_2-3}} \left( \Phi_1^{q_2}\bar{\Phi}^{q_1}_2 + \bar{\Phi}^{q_2}_1 \Phi_2^{q_1}\right)\ ,
\eeq
where $M_{PL}=2.4 \times 10^{18}$\,GeV denotes the reduced Planck scale.
When supersymmetry is spontaneously broken in a separate sector, the above superpotential contributes to the axion potential 
through the supergravity effects,%
\footnote{In supergravity, a superpotential term $W_i$ directly appears in the scalar potential as
\beq
V= (n_i - 3)m_{3/2}\times W_i + h.c.\ ,
\eeq 
with $n_i$ being the mass dimension of $W_i$.} 
\beq
\label{eq:explicitV}
V \sim \frac{1}{2} m_a^2 F_a^2 \left(\frac{a}{F_a}\right)^2 + \frac{m_{3/2}}{M_{\rm PL}^{q_1+q_2-3}} 
\Lambda_1^{q_2}\Lambda_2^{q_1}\frac{a}{F_a}+ h.c. + \cdots \ ,
\eeq
where $m_{3/2}$ denotes the gravitino mass.
In the final expression, we use $\Phi_1^{q_2}\bar{\Phi}^{q_1}_2 = \Lambda^{q_2 + q+1}/2^{(q_2+q_1)/2} e^{ia/F_a}$,
and the intrinsic $\theta$ angle of QCD is absorbed by the definition of the axion field.
The first term represents the axion mass term due to the QCD effects~\cite{Weinberg:1977ma}, 
\begin{eqnarray}
m_a^2 \simeq \frac{m_u m_d}{(m_u+m_d)^2}  \frac{m_\pi^2 f_\pi^2}{F_a^2}\ ,
\end{eqnarray}
where $m_{u,d}$ are the $u$- and $d$-quark masses, $m_\pi$ the pion mass, and $f_\pi \simeq 93$\,MeV the pion decay constant.

As a result, the effective $\theta$ angle at the vacuum of the axion is given by,
\beq
\theta_{\rm eff}&\simeq& \frac{1}{m_a^2 F_a^2}   \frac{m_{3/2}}{M_{\rm PL}^{q_1+q_2-3}} \Lambda_1^{q_2}\Lambda_2^{q_1}\nonumber\\
&\sim&  10^{66-6.4(q_1+q_2)}\times \left( \frac{m_{3/2}}{10^6\,{\rm GeV}}\right)^2 \left( \frac{0.08\,{\rm GeV}}{\sqrt{m_aF_a}}\right)^4
\left( \frac{\Lambda_1}{10^{12}\,{\rm GeV}}\right)^{q_2}  \left( \frac{\Lambda_2}{10^{12}\,{\rm GeV}}\right)^{q_1}.
\eeq
Thus, for $q_1+q_2 \gsim 12$, $m_{3/2} = {\cal O}(10^6)$\,GeV, and $\L_{1,2} ={\cal O}(10^{12})$\,GeV, the explicit breaking terms 
of the global PQ symmetries are small enough to be consistent with 
the measurement of the neutron EDM, i.e. $\theta_{\rm eff} < 10^{-11}$~\cite{Baker:2006ts}.
In this way, a {\it high quality} global PQ symmetry appears as an accidental symmetry in the gauged PQ mechanism.

\subsection{Domain Wall Problem}
\label{sec:domainwall}
Before closing this section, let us briefly discuss the domain wall problem.
The anomaly cancelation condition in Eq.\,(\ref{eq:anomfree}) is generically solved by,
\begin{eqnarray}
N_1 = N_{\rm GCD}\times q_2\ , \quad N_2 = N_{\rm GCD}\times q_1\ ,
\end{eqnarray}
where $N_{\rm GCD} \in {\mathbb N}$ is the greatest common devisor of $N_{1,2}$.
Then, the anomalous coupling in Eq.\,(\ref{eq:anomSF}) is rewritten by,
\begin{eqnarray}
\label{eq:anomSF3}
W_{\rm anom} = \frac{N_{\rm GCD}}{8\pi^2}
\frac{\sqrt{2} A}{F_a}
 \sum_l  W_l^{\a}W_{l\a}\ ,
\end{eqnarray}
and hence,
\beq
\label{eq:anomSM}
\mathcal{L}_{\rm anom}=\frac{N_{\rm GCD}}{32\pi^2}\frac{a}{F_a}\sum_{l} F^{l}_{\mu\nu}\tilde{F}^{l\mu\nu}\ .
\eeq
It should be noted that the anomalous coupling of the axion respects a discrete symmetry, ${\mathbb Z}_{N_{\rm GCD}}$,
\begin{eqnarray}
\frac{a}{F_a} \to \frac{a}{F_a} + \frac{2\pi k}{N_{\rm GCD}}\ ,  \quad k = 0,\,\cdots,\,  N_{\rm GCD}  - 1 \ .
\end{eqnarray}
for $N_{\rm GCD} > 1$. 

The ${\mathbb Z}_{N_{\rm GCD}}$ symmetry is eventually broken in the vacuum of the axion.
Thus, the model with $N_{\rm GCD} > 1$ suffers from the domain wall problem if the global PQ symmetry 
is broken after inflation since the average of the axion field value in each Hubble volume is randomly distributed.
To avoid the domain wall problem, spontaneous breaking of the global PQ symmetry is required to take place before inflation,
which in turn requires a rather small inflation scale to avoid the axion isocurvature problem (see, e.g. 
Ref.~\cite{Kawasaki:2013ae,Kawasaki:2018qwp}).

For $N_{\rm GCD} = 1$, on the other hand, there is no discrete symmetry which is broken by the 
VEV of the axion field.
Still, however, there can be domain wall problems when the global PQ breaking takes place after inflation.
To see this problem, let us remember that there can be various types of cosmic string configurations
formed at spontaneous symmetry breaking of the PQ symmetries.
For example, when both the gauged and the global PQ symmetries are broken spontaneously after inflation,
there can be cosmic string configurations in which either the phase of $\Phi_{1}$ or $\Phi_{2}$ takes $0-2\pi$ around configurations.
It should be noted that those configurations are the global strings and not the local string.
Thus, the string tensions diverge in the limit of infinite volume which is cut off by the Hubble volume. 
The local string, on the other hand, corresponds to the configurations in which the phases of $\Phi_{1}$ and $\Phi_2$ wind
$q_1$ times and $q_2$ times simultaneously. 
With the $U(1)_{gPQ}$ gauge field winding simultaneously, the tension of the local string is finite even in the limit of infinite volume for 
the local string.

A striking difference between the global strings and the local strings is how
the axion field winds around the strings. 
Around the local strings, only the would-be-Goldstone field winds, while the axion winds around the global strings.
Thus, when the axion potential is generated at around the QCD scale, the axion domain walls are formed 
only around the global strings, while they are not formed around the local strings.
Once the domain walls are formed around the global strings, they immediately dominate 
over the energy density of the universe, which causes the domain wall problem. 
Therefore, for the domain wall problems not to occur, the local strings should be formed preferentially at the phase transition.

The string tensions of the global strings and the local strings, however, depend on model parameters. 
Thus, there is no guarantee that only the local strings preferentially survive in the course of the 
cosmic evolution. 
As an example, let us consider a case with $\vev{\Phi_1} \gg \vev{\Phi_2}$.
In this case, the cosmic strings are formed at the first phase transition, i.e.  $\vev{\Phi_1}\neq 0$ with $\vev{\Phi_2} = 0$.
At this stage, strings around which the phase of $\Phi_1$ winds just once are
expected to be dominantly formed. They are {\it local} because we can take
an appropriate charge normalization for the $U(1)_{gPQ}$.
As the temperature of the universe decreases, the string networks follow the scaling solution 
where the number of the cosmic strings in each Hubble volume becomes constant (see, e.g., Ref.~\cite{Vilenkin:2000jqa}).

Once the temperature becomes lower than the scale of the second  phase transition, i.e., $\vev{\Phi_2}\neq 0$, 
the {\it local} strings formed at the first phase transition become no more the local strings.%
\footnote{The configuration of the gauge field formed at the first phase transition does not coincide with the one 
required for the local string with $\vev\Phi_2 \neq 0$.
}
Besides, formations of the global strings of  $\Phi_2$ are also expected at the second phase transition
in which the phase of $\Phi_2$ winds just once.
To form a genuine local string, it is required to bundle $q_1$ ex-local strings (formed by $\Phi_1$) and $q_2$ global strings (formed by $\Phi_2$)
into a single string.
However, the confluence of global strings into a local string is quite unlikely as there is no correlation between 
the nature of the cosmic strings in the adjacent Hubble volumes.
Therefore, when $\vev{\Phi_{1}} \gg \vev{\Phi_2}$, the domain wall problem is expected to be not avoidable 
even if $N_{\rm GCD} = 1$.%
\footnote{As there is no corresponding discrete symmetry, the domain wall is not stable completely.
For $\vev{\Phi_1}\gg \vev{\Phi_2}$, however, the decay rate (i.e., the puncture rate and/or the rate of the breaking off)
is highly suppressed.
}

In summary, let us list up possibilities to avoid the domain wall problem.
The first possibility is  a trivial one where both the gauged and the global PQ symmetries are broken before inflation.
This solution does not require $N_{\rm GCD}= 1$.
In this possibility, there is a constraint on the Hubble scale during inflation from the axion isocurvature problem. 

The next possibility is only applicable for $N_{\rm GCD} = 1$ with $q_1 = 1$ and $q_2 = N (>1)$.
Here, it is  assumed that the first phase transition (i.e. $\vev{\Phi_1}\neq 0$) takes place 
before inflation while the second phase transition (i.e. $\vev{\Phi_2} \neq 0$) occurs
after inflation.
In this second possibility, the {\it local} strings formed at the first phase transition are inflated away.
The global strings formed at the second phase transition, on the other hand, do not cause the domain wall problem 
as each of the global string is attached to only one domain wall~\cite{Hiramatsu:2010yn,Hiramatsu:2012gg}.

In addition to these two possibilities, there can be another possibility 
which is applicable for $N_{\rm GCD} = 1$ with $\vev{\Phi_1} \sim \vev{\Phi_2}$.
In this case, there can be a possibility where the local strings are preferentially formed at the phase transition.
Besides, the axion domain wall attached to the global strings may have very short lifetime for $N_{DW}  = 1$
even if they are formed.
To confirm this possibility, detailed numerical simulations are required, which goes beyond the scope of this paper. 

It should be noted that the second possibility (and the third possibility if numerically confirmed) is 
one of the advantages of the gauged PQ mechanism over the models in which 
the global PQ symmetry results from an exact discrete symmetry, such as ${\mathbb Z}_N$.
In such models, the axion potential also respects the ${\mathbb Z}_N$ symmetry, 
and hence, the domain wall problem is not avoidable when the global PQ symmetry is spontaneously 
broken after inflation.
In the gauged PQ models, on the other hand, it is possible that the global PQ symmetry is broken after inflation 
without causing the domain wall problem nor the axion isocurvature problem.

\section{Dynamical supersymmetry/PQ symmetry Breaking }
\label{sec:proto}
In this section, we discuss a model of a simultaneous breaking of supersymmetry and  the {\it global} PQ symmetry.
As we are interested in solutions to the strong $CP$-problem without severe fine-tuning, it is natural to seek models
in which the PQ breaking scale is generate by dynamical transmutation. 
Thus, in the following, we construct a model of a simultaneous supersymmetry/PQ symmetry breaking sector based on a strong dynamics.
For now, we do not consider the gauged PQ mechanism which will be implemented in the next section.
\subsection{Simultaneous Breaking of Supersymmetry and  Global PQ Symmetry}
As the simplest example of the dynamical supersymmetry breaking models, 
we consider a model of supersymmetry breaking based on $SU(2) $  gauge dynamics
(the IYIT model)~\cite{Izawa:1996pk, Intriligator:1996pu}.
The advantage of this model is that the nature of dynamical supersymmetry breaking is calculable by using 
effective composite states.

The model consists of four $SU(2) $  doublets, $Q_i~(i=1-4)$, and six singlets, $Z_{ij}=-Z_{ji}~(i,j=1-4)$. 
Those superfields couple via the superpotential
\beq
\label{eq:tree}
W_{IYIT}=\sum\lambda^{kl}_{ij}Z^{ij}Q_kQ_l
\eeq
where $\lambda^{kl}_{ij}$ denote coupling constants with $\lambda^{kl}_{ij}=-\lambda^{kl}_{ji}=-\lambda^{lk}_{ij}$. 
The maximal non-abelian global symmetry of the IYIT model is $SU(4)$ flavor symmetry, $SU(4)_f$, 
which is broken by $\lambda^{kl}_{ij}$.

The superpotential Eq.\,(\ref{eq:tree}) respects a global $U(1)_A$ symmetry with charges, $Z$'s$(+2)$, $Q$'s$(-1)$,
and a continuous $R$-symmetry, $U(1)_R$, with $Z$'s$(+2)$, $Q$'s$(0)$ (Tab.\,\ref{tab:IYIT}).
The former is broken down to the discrete subgroup, $\mathbb{Z}_4$, by the $SU(2) $ anomaly, while the latter is free from the $SU(2) $  anomaly.
As we seek a solution to the strong $CP$ problem not relying on global symmetries, we consider that the $\mathbb{Z}_4$
and $U(1)_R$ symmetries are accidental symmetries and are broken by Planck suppressed operators.

It should be noted, however, that a discrete 
subgroup of $U(1)_R$, $\mathbb{Z}_{NR}~(N>2)$, 
plays crucial roles
in constructing the SSM.
Without $\mathbb{Z}_{NR}~(N>2)$ symmetry, the VEV of the superpotential is expected to be of the order of the 
Planck scale.
Such a large VEV of the superpotential, in turn, does not allow a supersymmetry breaking scale lower than the Planck scale 
due to the condition for the flat present universe.
In addition, it is also known that $R$-symmetry (or at least an approximate $R$-symmetry) 
is relevant for supersymmetry breaking vacua to be stable\,\cite{Affleck:1984xz, Nelson:1993nf}.
Given its importance, we assume that the $\mathbb{Z}_{NR}~(N>2)$ symmetry is 
an exact discrete gauge symmetry~\cite{Krauss:1988zc,Preskill:1990bm,
Preskill:1991kd,Banks:1991xj,Ibanez:1991hv,Ibanez:1992ji,Csaki:1997aw}.%
\footnote{In Ref.\,\cite{Harigaya:2013vja}, it is proposed to achieve the global PQ symmetry
as an accidental symmetry protected by the exact discrete $R$-symmetry without relying on the gauged PQ mechanism.
}
In this paper, we take the simplest possibility, ${\mathbb Z}_{4R}$,
assuming a presence of an extra multiplet of the ${\bf 5}$,~$\bar{\bf 5}$ representations of the $SU(5)$ GUT.
The ${\mathbb Z}_{4R}$ symmetry is free from the Standard Model anomaly when the $R$-charges of the bilinear term 
of the Higgs doublets and that of the extra multiplets are vanishing~\cite{Kurosawa:2001iq,Lee:2010gv,Fallbacher:2011xg,Evans:2011mf}.%
\footnote{For GUT models which are consistent with the ${\mathbb Z}_{4R}$ symmetry, see, e.g.,~\cite{Izawa:1997he,Harigaya:2015zea}.}

In this model, we identify the global PQ symmetry with a $U(1)$ subgroup of $SU(4)_f$ 
(Tab.\,\ref{tab:IYIT}).
As it is a subgroup of $SU(4)_f$, the PQ symmetry  is free from the $SU(2) $  anomaly.
Under the global $U(1)_{PQ}$ symmetry, the superpotential is reduced to 
\beq
W_{IYIT} = \lambda_{12}^{12} Z^{12}Q_1 Q_2 + \lambda_{34}^{34} Z^{34}Q_3 Q_4 + \sum\tilde\lambda^{kl}_{ij}Z^{ij}Q_kQ_l
\eeq
where $\lambda$'s are dimensionless coupling constants with $\tilde\lambda^{kl}_{ij} = 0$
for $ij = 12$, $34$ or $kl$ = $12$, $34$. 
Hereafter, we take $\lambda^{12}_{12} = \lambda^{34}_{34} = \lambda$ for simplicity,%
although it is straightforward to extend the following analysis for $\lambda^{12}_{12}\neq  \lambda^{34}_{34}$.
As we will see shortly, the PQ symmetry is spontaneously broken by the VEV of $Q_{1}Q_2$ and $Q_3Q_4$.

\begin{table}[t]
\caption{\sl Charge assignment of the simultaneous symmetry breaking model. 
The chiral superfields, $Q$'s, and $Z$'s, are the $SU(2) $  doublets and singlets of the IYIT model, respectively.
The $U(1)_{PQ}$ symmetry is a subgroup of the maximum flavor symmetry of the IYIT model.
The KSVZ extra multiplets consist of the ${\bf 5}$ and $\bar {\bf 5}$ representations of the $SU(5)$ GUT group.
The U(1)$_R$ and $U(1)_A$ symmetries are accidental symmetries of the IYIT model.
A discrete subgroup of U(1)$_R$, i.e. ${\mathbb Z}_{4R}$ is assumed to be an exact symmetry.
The $R$-charges of the KSVZ extra multiplets are taken to be $r_5 + r_{\bar 5} = 2$.
} 
\small{
\begin{center}
\begin{tabular}{|c|c|c|c|c|c||c|c|}
\hline
 & $Q_{1,2}$ & $Q_{3,4}$ & $Z_{12} (Z_-)$ & $Z_{34} (Z_+)$ &  $Z_{13,14,23,24} (Z_{0}^a\,(a=1-4))$ & ${\bf 5}$ & $\bar{\bf 5}$ \\ \hline
$SU(2) $  & ${\bf 2}$ &${\bf 2}$ &${\bf 1}$ & ${\bf 1}$ & ${\bf 1}$  & ${\bf 1}$ &${\bf 1}$ \\ \hline
$U(1)_{PQ}$ & $+1$ &$-1$ &$-2$ & $+2$ & $0$  & $-2$ &$0$ \\ \hline \hline
$U(1)_{R}$ & $0$ &$0$ &$+2$ & $+2$ & $+2$ & $r_5$ & $r_{\bar 5}$   \\ \hline
$U(1)_A$ & $-1$ &$-1$ &$+2$ & $+2$ & $+2$  & $+2$ & $+0$   \\ \hline 
\end{tabular}
\end{center}}
\label{tab:IYIT}
\end{table}%

By assuming the KSVZ axion model, the PQ symmetry is communicated to the SSM sector
through couplings to the KSVZ extra multiplets in ${\bf 5}$ and $\bar{\bf 5}$ representations of the $SU(5)$ GUT,%
\footnote{The KSVZ extra multiplets should be distinguished the extra multiplets required to cancel the Standard Model anomaly
of the ${\mathbb Z}_{4R}$ symmetry.}
\beq
\label{eq:KSVZ1}
W = \frac{1}{M_{\rm PL}} Q_1 Q_2  {\bf 5}\,\bar{\bf 5} \ .
\eeq
The PQ charges of the KSVZ extra multiplets are given in Tab.\,\ref{tab:IYIT}.
Hereafter, we assume that there are $N_f$ flavors of the KSVZ extra multiplets.
Once $\vev{Q_{3}Q_4}$ spontaneously breaks the PQ symmetry,  the axion couples to the SSM sector via Eq.\,(\ref{eq:KSVZ1}) 
and the extra multiplets obtain masses of ${\cal O}(\vev{Q_3Q_4}/M_{PL})$.

Now, let us discuss how supersymmetry and the PQ symmetry are  broken spontaneously.
Below the dynamical scale of $SU(2) $  dynamics, $\Lambda$,  the IYIT model is well described by using the composite fields, 
$M_{ij} \sim Q_iQ_j$, with an effective superpotential,
\beq
\label{eq:IYIT}
W_{\rm eff} \sim \lambda \L Z_- {M}_+ +  \lambda \L Z_+ {M}_- + \tilde\lambda_{ab} \L Z_0^a {M}_0^b +{\cal X}(2 {M}_+ {M}_- + {M}_0^a {M}_0^a - \Lambda^2)\ .
\eeq
Here, ${M}_{+} \sim Q_1Q_2/\L$ and ${M}_{-} \sim Q_3Q_4/\L$ denote the PQ charged mesons,
while ${M}_{0}^a$ $(a= 1-4)$ are the PQ neutral mesons. 
The coupling constants $\tilde \lambda$ and the singlets $Z_0$'s are also rearranged accordingly.
In the effective superpotential, the quantum modified constraint $2{M}_+ {M}_- + {M}_0 {M}_0 - \Lambda^2=0$~\cite{Seiberg:1994bz} 
is implemented by a Lagrange multiplier field $\cal X$.

By assuming that $\lambda$'s are perturbative, and $\lambda_{\pm}(=\lambda)$ are smaller than $\tilde\lambda$'s, 
the VEVs of $M_\pm$ are given by
\begin{eqnarray}
\label{eq:vacuum}
\langle M_+ \rangle= \frac{1}{\sqrt 2} \Lambda ,
\quad
\langle  M_-\rangle  = \frac{1}{\sqrt 2} \Lambda\ .
\end{eqnarray}
Other fields do not obtain VEVs of ${\cal O}(\Lambda)$.%
\footnote{The scalar components of $Z_\pm$ and ${\cal X}$ obtain small VEVs of ${\cal O}(m_{3/2})$. }
At this vacuum, the PQ symmetry is spontaneously broken by $\langle M_\pm \rangle$ while 
supersymmetry is broken by the VEVs of the $F$-components of $Z_{\pm}$, i.e.,
\begin{eqnarray}
\label{eq:FZ}
F_{Z_\pm} \sim \frac{1}{\sqrt 2} \lambda \Lambda^2 \ ,
\end{eqnarray}
simultaneously.

Here, let us comment that the ${\mathbb Z}_{4R}$ is not enough to restrict the superpotential in the form of Eq.\,(\ref{eq:tree}).
In fact, there can be superpotential terms such as $Z_0^3$ or $Z_0 Z_+ Z_-$ without the $U(1)_A$ (or ${\mathbb Z}_4$) symmetry.
As those terms make the supersymmetry breaking vacuum in Eqs.\,(\ref{eq:vacuum}) and (\ref{eq:FZ}) metastable, the coefficients 
of those terms should be rather suppressed to make the vacuum long lived.
Such suppression can be achieved, for example, by assuming that a subgroup of ${\mathbb Z}_4$ and $U(1)_{PQ}$ is an exact symmetry
where $Z_{0}$'s are charged but $Z_\pm$ are neutral.%
\footnote{As this symmetry is not broken spontaneously at the vacuum, and hence, $Z_0$'s and $M_0$'s are predicted to be stable.
Thus, the simultaneous breaking of the IYIT sector should take place before inflation to avoid the production of those stable particles
if we assume the above symmetry.}
It is also possible to suppress the unwanted terms by extending the SU(2) dynamics of the IYIT sector into a conformal window by adding extra doublets~\cite{Ibe:2005pj,Ibe:2005qv,Ibe:2007wp}.

\subsection{Axion Supermultiplet}
The degeneracy due to the PQ symmetry breaking is parametrized by the axion superfield $A$,
\begin{eqnarray}
\label{eq:axionSF}
 M_+ =  \frac{1}{\sqrt 2}\Lambda e^{A/\Lambda} \ , \quad
 M_-=  \frac{1}{\sqrt 2}\Lambda e^{-A/\Lambda} \ , 
\end{eqnarray}
with which the PQ symmetry is realized by 
\begin{eqnarray}
A/\Lambda \to A/\Lambda + i \alpha \ , \quad (\alpha = 0 - 2\pi)\ .
\end{eqnarray}
Here, we reduce the domain of the $U(1)_{PQ}$ rotation parameter from $\alpha = 0-4\pi$ to $\alpha = 0-2\pi$, 
since all the $SU(2) $  gauge invariant fields have the PQ charge of $\pm 2$ (see Tab.\,\ref{tab:IYIT}).
In other words, the sign changes of $Q$'s by a phase rotation with $\alpha = 2\pi$ 
can be absorbed by a part of $SU(2) $  transformation.

The effective K\"ahler potential and superpotential of $M_\pm$ and $Z_\pm$ are given by,
\begin{eqnarray}
\label{eq:axionEFF}
K_{\rm eff} &\sim& |Z_+|^2 + |Z_-|^2 + | M_+|^2 + | M_+|^2 + \cdots\ , \\
W_{\rm eff} &\sim& \lambda 
\L M_+ Z_- + \lambda \L M_- Z_+ \ ,
\end{eqnarray}
where the ellipses denote the higher dimensional operators.
By substituting the axion superfield, the effective theory is reduced to
\begin{eqnarray}
K_{\rm eff} &\sim& | Z_+|^2 + | Z_-|^2 + \frac{1}{2}(A^\dagger + A)^2 +\cdots  \ , \\
\label{eq:axionEFF2}
W_{\rm eff} &\sim& \frac{1}{\sqrt 2}\lambda \Lambda^2 (Z_- e^{A/\L} + Z_+e^{-A/\L}) \ ,
\end{eqnarray}
with some irrelevant holomorphic terms omitted in the K\"ahler potential.
The scalar potential is accordingly given by,%
\footnote{Throughout the paper, we use the same symbols to describe the superfields and their scalar components.}
\begin{eqnarray}
V &\sim& \l^2 \L^4 \cosh\left(\frac{A^\dagger + A }{\L}\right) 
+ \frac{1}{2}\l^2 \L^4\left| Z_+ e^{-A/\L} - Z_- e^{A/\L}\right|^2 \\
&\sim& \l^2 \L^4 \cosh\left(\frac{A^\dagger + A }{\L}\right) 
+ \l^2 \L^4|T|^2 \ .
\end{eqnarray}
In the final expression, we rearranged the scalar fields by introducing complex scalar fields $S$ and $T$,
\begin{eqnarray}
Z_+ &=&  \frac{1}{\sqrt{2}}(S+T)e^{-A/\L}\ , \\
Z_- &=&  \frac{1}{\sqrt{2}}(S-T)e^{A/\L}\ , 
\end{eqnarray}
so that the PQ symmetry is manifest in the scalar potential.

The above scalar potential shows that the complex scalar $T$ and the real component of $A$ (the saxion)
obtain masses of $\lambda\Lambda$, around their origins.
The complex scalar filed $S$ (the pseudo-flat direction) and the imaginary part of $A$ (the axion $a$), on the other hand, remain massless.
The pseudo-flat direction eventually obtains a mass from the higher order terms in the K\"ahler potential. 
For perturbative $\lambda$  and $\tilde\lambda$, the mass is dominated by the one-loop contributions 
~\cite{Chacko:1998si,Ibe:2009dx,Ibe:2010ym},
\begin{eqnarray}
m_S^2 \simeq \frac{1}{32\pi^2} \left({\lambda^2 (2\log2-1)}
+ \frac{4\lambda^4}{3\tilde\lambda^2}  \right)
\frac{F_S^2}{ \Lambda^2}\ , \quad (F_S = \lambda \Lambda^2)\ ,
\end{eqnarray}
with which the pseudo-flat direction is stabilized at its origin.%
\footnote{Here, we neglect the one-loop contributions from the $U(1)_{gPQ}$ gauge interaction 
by assuming that the gauge coupling constant is small.
The contributions from the gauge interaction, in fact, destabilize the origin of the pseudo flat direction~\cite{Dine:2006xt,Ibe:2009dx,Ibe:2010ym}.
}

The superpotential in Eq.\,(\ref{eq:axionEFF2}) also shows that the fermion partners of $A$ (the  axino) 
and $T$ obtain a Dirac mass of $\lambda \Lambda$, with each other.
The fermion partner of $S$ corresponds to the goldstino which is absorbed into the gravitino by the super-Higgs mechanism.

Putting together, the model achieves dynamical breaking of supersymmetry and the PQ breaking simultaneously. 
The axion supermultiplet splits into a massless axion and massive saxion/axino with masses of the supersymmetry/PQ breaking scale.
The axion couples to the SSM sector via the coupling in Eq.\,(\ref{eq:KSVZ1}), i.e.,
\begin{eqnarray}
W 
\sim \frac{\L^2}{\sqrt{2}M_{\rm PL}} e^{A/{\Lambda} } \,\bar{\mathbf 5}\,{\mathbf 5} 
\sim \frac{\L^2}{\sqrt{2}M_{\rm PL}} e^{{ia}/{f_a} } \,\bar{\mathbf 5}\,{\mathbf 5} \ ,
\end{eqnarray}
where $a = \sqrt{2} \operatorname{Im}[A]$ denotes the axion field and  $f_a = \sqrt{2}\Lambda$.
After integrating out the extra KSVZ multiplets, the axion couples to the SM gauge fields through
\begin{eqnarray}
\label{eq:anomSF2}
W_{\rm anom} = \frac{N_f}{8\pi^2}
\frac{A}{\Lambda}
\sum_l W_l^{\a}W_{l\a}\ ,
\end{eqnarray}
with which the strong $CP$ problem is solved  

\subsection{Explicit Breaking of the PQ symmetry}
Now, let us discuss explicit breaking of the global PQ symmetry expected in quantum gravity.
In this model, the most relevant terms which break the global PQ symmetry are given by,
\footnote{Here, we require that $U(1)_{PQ}$ is not broken by renormalizable interactions as a part of definition of the global symmetry.} 
\beq
W \sim \frac{\k}{M_{\rm PL}^2} 
Z_{+} (Q_{1}Q_{2})^2
+\frac{\k}{M_{\rm PL}^2}  Z_-(Q_3 Q_4)^2
\sim \frac{\k\L^2}{M_{\rm PL}^2} Z_{\pm} {M}_{\pm}^2\ .
\eeq
with $\k$ being a dimensionless coupling constant.%
\footnote{Lower dimensional operators which break the PQ symmetry, such as $Z_+^4/M_{PL}$, 
are forbidden by the ${\mathbb Z}_{4R}$ symmetry.
} 
The corresponding symmetry breaking terms in the scalar potentials are given by,
\beq
V\sim \frac{1}{2}m_a^2 f_a^2\left(\frac{a}{f_a}\right)^2 + \lambda\kappa\left(\frac{\Lambda}{M_{\rm PL}}\right)^2 \Lambda^4 e^{i \frac{a}{f_a}} + h.c.\ .
\eeq 
Here, we inserted the VEVs of $M_\pm$ and those of $F$-terms of $Z_\pm$.
Due to the explicit breaking, the VEV of the axion, and hence, the effective $\theta$ angle is shifted to
\beq
\label{eq:quality}
\theta_{\rm eff} = \frac{\vev{a}}{f_a}&\simeq&\left. 
\operatorname{Im}[\k\l] \left(\frac{\Lambda}{M_{\rm PL}}\right)^2 \left(\frac{\Lambda^4}{m_a^2 f_a^2}\right)\right|_{\text{\scalebox{1.2}{mod 2$\pi$}}} \\
&\simeq&\left.10^{40} \times \operatorname{Im}[\k\l] 
\left(\frac{0.08\,{\rm GeV}}{\sqrt{m_a f_a}}\right)^4 \left( \frac{\Lambda}{10^{12}\,{\rm GeV}}\right)^6 \right|_{\text{\scalebox{1.2}{mod 2$\pi$}}}  \ .
\eeq
Thus, unless $\operatorname{Im}[\k\l]$ is finely tuned to be smaller than ${\cal O}(10^{-11})$, the effective $\theta$ angle is too large to 
be consistent with the measurement of the neutron electric dipole moment (EDM)~\cite{Baker:2006ts}.

\section{Gauged PQ Extension of simultaneous breaking model}
\label{sec:model}
Let us now implement the gauged PQ mechanism to the model of the simultaneous breaking of supersymmetry and the PQ symmetry in section\,\ref{sec:proto}.
For that purpose, we introduce an additional sector based on $SU(3)$ dynamics which breaks a PQ symmetry spontaneously.
In the following, we call this model the $SU(3)'$ model, and put primes on the superfields and the symmetry groups
in this sector.
\subsection{$SU(3)'$ PQ Symmetry Breaking Model}
\label{sec:pqbreaking}
The $SU(3)'$ model consists of three flavors of the (anti-)fundamental representation of $SU(3)$, $Q'$, $\bar{Q}'$, 
and nine $SU(3)$ singlets, $Z'$.
The charge assignment of the global symmetries is given in Tab.\,\ref{tab:23}. 
Under these symmetries, they couple via the superpotential
\beq
W_{PQ}=\lambda'^{kl}_{ij}Z^{'ij}Q'_k\bar{Q}'_l\ ,
\eeq
where $\lambda'^{kl}_{ij}$ denote coupling constants with $(i,j,k,l=1-3)$. 
The baryon symmetry, $U(1)_B$, is identified with the global PQ symmetry, $U(1)'_{PQ}$,
while the maximal flavor symmetry, $SU(3)_L\times SU(3)_R$, is completely broken by 
$\lambda'$'s.

In addition to the global $U(1)'_{PQ}$ symmetry, the superpotential possesses a continuous $R$-symmetry
and a $U(1)_A'$ symmetry (broken down to a ${\mathbb Z}_6$ symmetry by the $SU(3)'$ anomaly) in Tab.\,\ref{tab:23}. 
As discussed previously, however, we consider that only ${\mathbb Z}_{4R}$ is an exact symmetry,
and assume that $U(1)_R$ and $U(1)_A'$ are accidental symmetries broken by higher dimensional operators.%
\footnote{Without $U(1)_A'$ (or ${\mathbb Z}_6$), the superpotential terms such as $Z'^3$ are allowed
even if we assume the ${\mathbb Z}_{4R}$ symmetry.
Such terms, however,  do not change the following discussion.}

\begin{table}[t]
\caption{\sl Charge assignment of the dynamical PQ symmetry breaking sector.
The chiral superfields, $Q'$'s, and $Z'$'s, are the $SU(3)'$ triplets and singlets, respectively.
The $U(1)'_{PQ}$ symmetry corresponds to $U(1)_B$ symmetry in the $SU(3)'$ sector.
The KSVZ extra multiplets are denoted by ${\bf 5}'$ and $\bar {\bf 5}'$.
The U(1)$_R$ and $U(1)_A'$ symmetries are accidental symmetries,
with ${\mathbb Z}_{4R}$ being an exact symmetry.
The $R$-charges of the KSVZ extra multiplets are taken to be $r_5' + r_{\bar 5}' = 2$.
} 
{
\begin{center}
\begin{tabular}{|c|Y|Y|Y||Y|Y|}
\hline
 & $Q' $ &$\bar{Q}'$  &$Z'$ &${\bf 5}' $ & $\bar{ {\bf 5}}' $  \\ \hline
 $SU(3)'$ & $\bf{3}$ &$\bar{\bf{3}}$ &$\bf{1}$ &$\bf{1}$ &$\bf{1}$  \\ \hline
$U(1)'_{PQ}$ & $+1$ &$-1$ &$0$ &$3$ & $0$   \\ \hline \hline
$U(1)_R$ & $0$ &$0$ &$+2$ &$r_5'$ & $r_{\bar 5}'$    \\ \hline
$U(1)_A'$ & $+1$ &$+1$ &$-2$ &$-3$ & $0$   \\ \hline 
\end{tabular}
\end{center}}
\label{tab:23}
\end{table}%

Below the dynamical scale of $SU(3)'$, $\L'$, the $SU(3)'$ sector is well described by the composite mesons and baryons,
\begin{eqnarray}
M' \sim Q'\bar{Q}'/\L'\ , \quad B_+' \sim Q'Q'Q'/\L'^2\ , \quad {B}_-' \sim \bar{Q}'\bar{Q}'\bar{Q}'/\L'^2\ ,
\end{eqnarray}
with an effective superpotential,
\beq
\label{eq:effSU3}
W'_{\rm eff}&=&\lambda' \Lambda' Z' M' + {\cal X}' (B'_+ B'_{-}+{\rm det}(M')/\Lambda'-\Lambda^{'2})\ . 
\eeq
Here, the second term implements the deformed moduli constraint by a  Lagrange multiplier  field $\cal X'$~\cite{Seiberg:1994bz}.
The mesons are neutral under $U(1)_{PQ}'$ while the baryons have charges $\pm 3$.

From the superpotential in Eq.\,(\ref{eq:effSU3}), we find that the PQ symmetry is spontaneously broken  by the VEVs of 
$B_\pm$.
Accordingly, the vacuum is parametrized by the Goldstone superfield $A'$,%
\footnote{The origin of $A'$ is set at which $B'_+  = B'_-$, and $\vev{B'_+} \neq \vev{B'_-}$ for $\vev{A'} \neq 0$, accordingly.}
\begin{eqnarray}
\label{eq:axionSF2}
B_+' &=&\frac{1}{\sqrt 2} \L_2 e^{A'/\Lambda_2}\ , \\
B_-' &=& \frac{1}{\sqrt 2}\L_2 e^{-A'/\Lambda_2}\  ,
\end{eqnarray}
with $\L_2 = 
\sqrt{2}\L' $. 
By using $A'$, the PQ symmetry is non-linearly realized by 
\begin{eqnarray}
\frac{A'}{\Lambda_2} \to  \frac{A'}{\Lambda_2} + i\alpha' \ , \quad (\alpha' = 0 - 2\pi)\ .
\end{eqnarray}
As in the case of the IYIT sector, the domain of the PQ symmetry is reduced from $\alpha' = 0 - 6\pi$ to $\alpha' = 0 - 2\pi$
as the $SU(3)'$ invariant fields have the PQ charges of $\pm 3$.

The $U(1)'_{PQ}$ symmetry in this sector is also communicated to the SSM sector through the couplings to $N_f'$ flavors of the KSVZ extra multiplets, 
$\mathbf 5'$ and $\bar{\mathbf 5}'$. 
With the charge assignment in Tab.\,\ref{tab:23}, the baryons couple to the extra multiplets in the superpotential,
\beq
\label{eq:KSVZ2}
W =\frac{1}
{M_{\rm PL}^2} 
\bar{Q}'\bar{Q}'\bar{Q}'
{\bf 5}'\bar{\bf 5}' \sim \frac{\L'^2}{M_{\rm PL}^2} B_-'{\bf 5}'\bar{\bf 5}'\ .
\eeq
Once $U(1)'_{PQ}$ is broken, the axion obtains the anomalous coupling to the SSM gauge fields,
while the extra multiplets obtain masses of ${\cal O}(\Lambda'^3/M_{\rm PL}^2)$.

\subsection{Gauged PQ Symmetry}
\begin{table}[t]
\caption{The charge assignment of the gauged PQ symmetry and the ${\mathbb Z}_{4R}$ symmetry.
The singlet fields $Y$'s and $Y'$'s are introduced to 
cancel the  self-triangle and gravitational anomalies of $U(1)_{gPQ}$ (see subsection~\ref{sec:anomcancel}). 
}
\small{
\begin{center}
\begin{tabular}{|c||Y|Y|Y|Y|Y|Y|Y|Y|Y|Y|}
\hline
 & $Z_\pm$&$M_\pm$ & $B'_+$&${B}_-'$ &${\bf 5}$ & $\bar{ {\bf 5}}$ &${\bf 5'}$ & $\bar{ {\bf 5}'}$ & $Y$ & $Y'$
  \\ \hline
$U(1)_{gPQ}$ & $\pm q_1$ &$\pm q_1$ &$-q_2$ &$q_2$ &$-q_1$  & $0$ &$q_2$ & $0$ & $q_1$ & $-q_2$
\\ \hline
${\mathbb Z}_{4R}$ & $2$ &$0$ &$0$ &$0$ &$r_5$  & $r_{\bar 5}$ &$r_5'$ & $r_{\bar 5}'$ & $1$ & $1$
\\ \hline
\end{tabular}
\end{center}}
\label{tab:gPQ}
\end{table}%
Now, we are ready to find out a model of the gauged PQ symmetry by combining 
the simultaneous supersymmetry and the PQ symmetry breaking model in section\,\ref{sec:proto}
and the PQ symmetry breaking model in subsection\,\ref{sec:pqbreaking}.
To apply the prescription in section\,\ref{sec:prescription}, let us first identify 
$\Phi_1$ with the meson operator $M_+$ in section\,\ref{sec:proto}
and $\Phi_2$ with the baryon operator $B'_+$, i.e.,
\begin{eqnarray}
\Phi_1 = M_+\  , \\
\bar\Phi_2\  = B_-' \ , 
\end{eqnarray}
and assign $U(1)_{gPQ}$ charges of $q_1$ and $-q_2$ to them~(Tab.\,\ref{tab:gPQ}).%
\footnote{The $U(1)_{gPQ}$ charges of $Q_{1,2}$ and $\bar{Q}'$'s corresponds to $q_1/2$ and $ q_2/3$, respectively.}
Then, the anomaly-free condition of the $U(1)_{gPQ}$ symmetry in Eq.\,(\ref{eq:anomfree}) is given by,
\begin{eqnarray}
\label{eq:anomfree2}
q_1 N_f - q_2 N_f'  = 0 \ .
\end{eqnarray}

Once the two sectors are put together by the gauged PQ symmetry, 
spontaneous breaking of the PQ symmetries in the two sectors
lead to the would-be Goldstone and the axion superfield.
The would-be Goldstone is absorbed into the massive $U(1)_{gPQ}$ gauge multiplet,
and the saxion and the axino in the axion supermultiplet obtain masses of the order of the supersymmetry breaking scale.
As a result, the simultaneous breaking model with the gauged PQ mechanism 
leaves only a light axion which couples to the Standard Model gauge fields via Eq.\,(\ref{eq:anomSM}).

\subsection{Accidental Global PQ Symmetry}
As discussed in the previous section, the global PQ symmetry can be explicitly broken by 
the $U(1)_{gPQ}$ invariant operator consisting of the fields in the two sectors.
Among the explicit breaking terms, the most relevant ones are given by,%
\footnote{
There are lower dimensional operators which break the global PQ symmetry with $M_\pm$ replaced by $M_{PL} \times Z_{\pm}$
in Eq.\,(\ref{eq:explicit}).
The explicit breaking effects of those operators are comparable to the ones of Eq.\,(\ref{eq:explicit})
due to suppressed $A$-term VEVs of $Z_\pm = {\cal O}(m_{3/2})$.}
\beq
\label{eq:explicit}
W 
&\sim& \frac{\k}{M_{\rm PL}^{3q_1+2q_2-4}} \left(Z_+(Q_{1}Q_2)^{q_2-1} (\bar{Q}'\bar{Q}'\bar{Q}')^{q_1} + 
Z_+(Q_{3}Q_4)^{q_2-1} (Q'Q'Q')^{q_1} 
\right)\ , \\
&\sim& \frac{\k\L^{q_2-1}\L'^{2 q_1} }{M_{\rm PL}^{3q_1+2q_2-4}} \left(Z_+ M_+^{q_2-1} {B}_-'^{q_1} + 
Z_- M_-^{q_2-1} {B}_+'^{q_1} \right)\ , 
\eeq
with $\k$ being a dimensionless coupling constant.
It should be noted that these terms are consistent with the ${\mathbb Z}_{4R}$ symmetry, 
and hence, no factor of $m_{3/2}$ is required unlike the terms in Eq.\,(\ref{eq:explicitW}).
These operators roughly contribute to the axion potential,  
\beq
V \sim \frac{1}{2} m_a^2 F_a^2 \left(\frac{a}{F_a}\right)^2 + \frac{\operatorname{Im}[\k \l]}{M_{\rm PL}^{3q_1+2q_2-4}} 
\Lambda^{2q_2}\Lambda'^{3q_1}\frac{a}{F_a}+ h.c. + \cdots\ ,
\eeq
where the VEVs of $M_\pm$,  $B_\pm'$, and those of the $F$-terms of $Z_{\pm}$ are inserted,
Therefore, in the simultaneous breaking model with the gauged PQ mechanism, the effective $\theta$ angle at the vacuum is given by,
\beq
\label{eq:theta3}
\theta_{\rm eff}&\simeq& \frac{1}{m_a^2 F_a^2}   \frac{\Lambda^{2q_2}\Lambda'^{3q_1}}{M_{\rm PL}^{3q_1+2q_2-4}} 
\\
&\sim&  10^{77.5-6.4(3q_1+2q_2)}\times\left( \frac{0.08\,{\rm GeV}}{\sqrt{m_aF_a}}\right)^4
\left( \frac{\Lambda}{10^{12}\,{\rm GeV}}\right)^{2q_2}  \left( \frac{\Lambda'}{10^{12}\,{\rm GeV}}\right)^{3q_1}\ . 
\eeq
Thus, for $3q_1+2 q_2 \gsim 14$, the explicit breaking of the global PQ symmetries are small enough to be consistent with 
the measurement of the neutron EDM, i.e., $\theta_{\rm eff}<10^{-11}$~\cite{Baker:2006ts}.

\subsection{Mass Spectrum of the KSVZ Multiplets}
\label{sec:mass}
The KSVZ multiplets, $(\bar{\mathbf 5},\bar{\mathbf 5})$ and $(\bar{\mathbf 5}',\bar{\mathbf 5}')$ were introduced to communicate the PQ symmetries to the SSM sector.
After PQ symmetry breaking, those extra multiplets obtain supersymmetric masses of  the order  of
\begin{eqnarray}
m_{KSVZ} &\sim& \frac{\Lambda^2}{M_{\rm PL}} \ , \\
\label{eq:KSVZp}
m_{KSVZ}' &\sim& \frac{\Lambda'^3}{M_{\rm PL}^2} \ , 
\end{eqnarray}
respectively (see Eqs.\,(\ref{eq:KSVZ1}) and (\ref{eq:KSVZ2})).
The scalar components of the KSVZ multiplets also obtain masses of ${\cal O}(m_{3/2})$ through supergravity effects.
Thus, most of the KSVZ multiplets become heavy and beyond the reach of the LHC experiments except for 
the fermion components of  $(\bar{\mathbf 5}',\bar{\mathbf 5}')$.%
\footnote{The extra multiplet to achieve the ${\mathbb Z}_{4R}$ symmetry also obtains the mass of ${\cal O}(m_{3/2})$
from the $R$-symmetry breaking effects~\cite{Casas:1992mk}.}

The KSVZ extra multiplets are assumed to couple to the SSM particle via,
\begin{eqnarray}
W 
&\sim&
 \frac{\e}{M_{\rm PL}} Q_1 Q_2 {\mathbf 5}\,  {\bar{\mathbf 5}}_{SM} + 
 \frac{\e'}
{M_{\rm PL}^2} 
\bar{Q}'\bar{Q}'\bar{Q}'
{\mathbf 5}' \,{\bar{\mathbf 5}}_{SM}  \ ,\\
&\sim&
 \epsilon \,m_{KSVZ} \,{\mathbf 5}\,  {\bar{\mathbf 5}}_{SM} + 
\epsilon'\, m_{KSVZ}' \,{\mathbf 5}' \,{\bar{\mathbf 5}}_{SM}  \ ,
\end{eqnarray}
where ${\bar{\mathbf 5}}_{SM}$  denotes the SSM matter multiplet, and $\e^{(')}$ are coefficients.
Here, we take $r_5 = r_5' = 1$ so that $\bar{\mathbf 5}$  and $\bar{\mathbf 5}'$  have 
the same $R$-charges with $\bar{\mathbf 5}_{SM}$.
Through the mixing terms, the KSVZ extra multiplets decay immediately into the SSM particles.

Finally, let us note that there can be mixing terms between (${\mathbf 5}$, $\bar{\mathbf 5}$) and 
(${\bar{\mathbf 5}'}$, ${\bar{\mathbf 5}'}$) through, 
\begin{eqnarray}
W \sim \frac{1}{M_{\rm PL}} Q_1 Q_2 {\mathbf 5}\,  {\bar{\mathbf 5}'} + 
 \frac{1}
{M_{\rm PL}^2} 
\bar{Q}'\bar{Q}'\bar{Q}'
{\mathbf 5}' \,{\bar{\mathbf 5}} \ .
\end{eqnarray}
Although these operators consist of the fields in the two PQ symmetric sectors,
they are invariant under not only the gauged PQ symmetry but also under the global PQ symmetries.
Thus, these terms do not affect $\theta_{\rm eff}$.
They do not affect the KSVZ mass spectrum significantly neither. 
From these reasons, we neglect these mixing terms throughout this paper.

\subsection{PQ Charges in the $SU(3)'$ Model}
For a given $q_1$ and $q_2$, there are upper limits on $\Lambda$  and $\Lambda'$ to achive
 a high-quality global PQ symmetry (see Eq.\,(\ref{eq:theta3})). 
The dynamical scales are also constrained from below for an appropriate supersymmetry breaking scale 
and for heavy enough KSVZ extra multiplets.
As a lower limit on the supersymmetry breaking scale, i.e., $\Lambda$, we require
\begin{eqnarray}
\label{eq:10TeV}
m_{3/2} &\simeq& \frac{\lambda\Lambda^2}{\sqrt{3}M_{\rm PL}} \gtrsim 10\,{\rm TeV}\ , 
\end{eqnarray}
so that the observed Higgs boson mass, $m_H \simeq 125$\,GeV, is achieved by 
the gravity mediated sfermion masses of ${\cal O}(m_{3/2})$.
As a lower limit on the KSVZ extra multiplets, we put
\begin{eqnarray}
\label{eq:750GeV}
m_{KSVZ}' &\simeq& \frac{\Lambda'^3}{M_{\rm PL}} \gtrsim 750\,{\rm GeV}\ ,
\end{eqnarray}
from the null results of the searches for a heavy $b$-type quark at the LHC experiments~\cite{Aad:2015kqa, Aad:2015gdg, Aad:2014efa, Aad:2015mba}.

\begin{figure}[t]
\begin{center}
\begin{minipage}{.7\linewidth}
  \includegraphics[width=\linewidth]{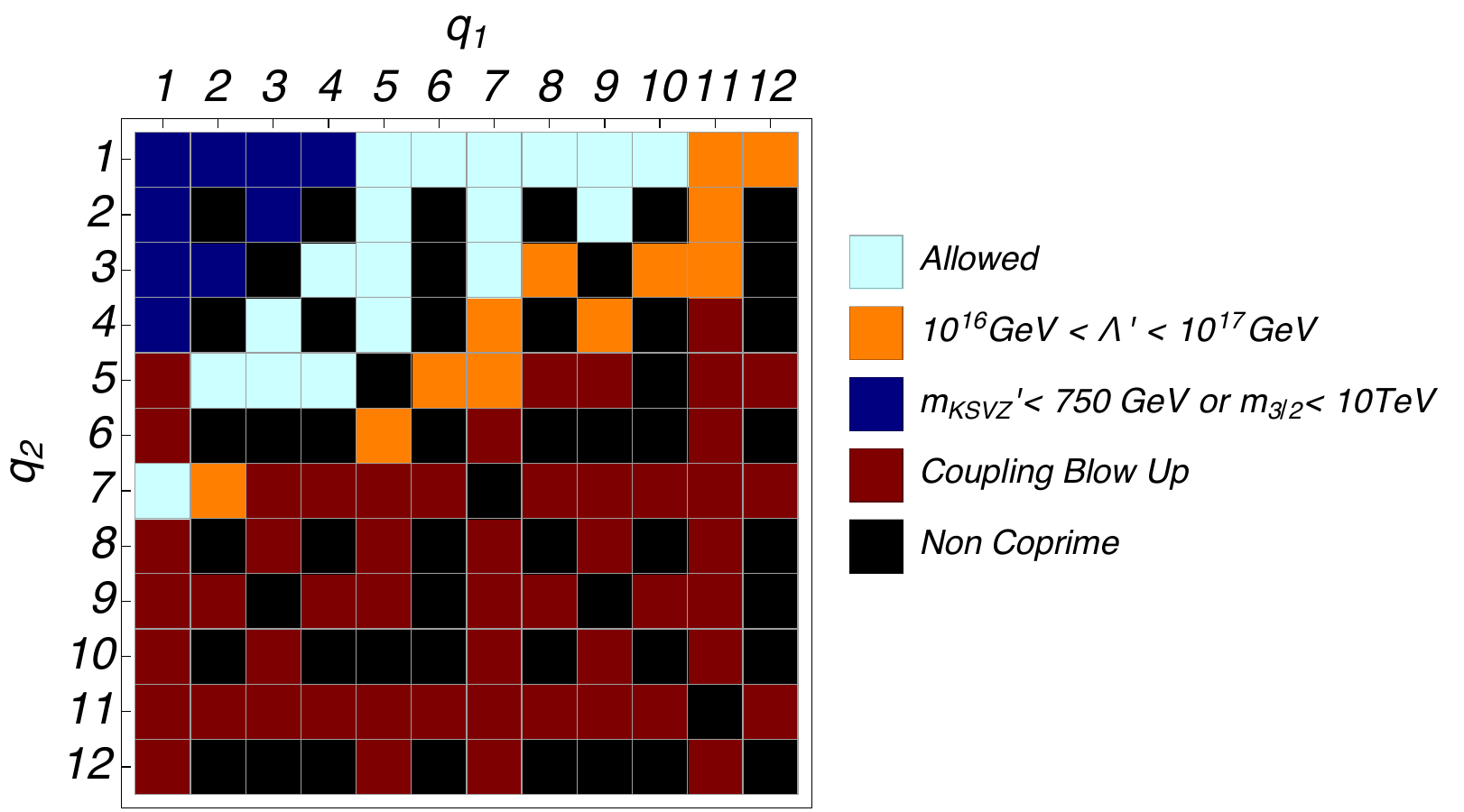}
 \end{minipage}
  \end{center}
\caption{\sl \small  Constraints on charges $q_1$ and $q_2$ in the $SU(3)'$ model for $N_{\rm GCD} = 1$.
The allowed charges are colored by light blue and orange, although the orange colored charges are allowed only for $\Lambda' \gtrsim M_{\rm GUT}$.
The charges colored by  blue lead to  too large $\theta_{\rm eff}$ or too light KSVZ extra multiplets.
The gauge coupling constants of the SSM blow up below the GUT scale for the charges colored by  red.
The black colored charges are excluded as they are not relatively prime.
}
\label{fig:const1}
\end{figure}

In Fig.\,\ref{fig:const1}, we show the charge choices for  $SU(3)'$ model for $N_{\rm eff} = 1$ and $\lambda= 1$.
The charges colored by blue are excluded, with which $\theta_{\rm eff}$ cannot be suppressed enough 
for $m_{3/2}\gtrsim {\cal O}(1)\,$TeV and $m_{\rm KSVZ}' \gtrsim 750$\,GeV.%
\footnote{The charges colored by  blue are not changed even if the lower limits on $m_{3/2}$ and $m_{KSVZ}'$ are relaxed 
to $m_{3/2}\gtrsim 100$\,GeV and $m_{KSVZ}' \gtrsim 100$\,GeV.}
In the figure, we require $\theta_{\rm eff} \lesssim 10^{-10}$ given ${\cal O}(1)$ uncertainties of the coefficients of the
explicit breaking terms.
The figure shows that these constraints exclude relatively small charges as the suppression of the explicit breaking term
relies on large PQ charges.

The perturbative coupling unification of the SSM gauge coupling constants also puts constraints on the charges.
From the anomaly-free condition in Eq.(\ref{eq:anomfree2}), $N_f$ and $N_f'$ are given by
\begin{eqnarray}
N_{f} = N_{\rm GCD}\times q_2\ , \quad N_{f}' = N_{\rm GCD} \times q_1\ .
\end{eqnarray}
As the extra multiplets contribute to the renormalization group evolutions of the SSM gauge coupling constants
and make them asymptotically non-free, the perturbative unification puts upper limits on $N_f$ and $N_f'$,
and hence, on $q_1$  and $q_2$.

In Fig.\,\ref{fig:const1}, we color the charges by red, with which $\theta_{\rm eff} \lesssim 10^{-10}$ is not compatible 
with the perturbative unification.
Here, we use the renormalization group equation at the one-loop level and require that $g_{1,2,3} < 4\pi$
below the GUT scale, i.e., $M_{\rm GUT} \simeq 10^{16}$\,GeV.
We also take the masses of the sfermions, the heavy charged/neutral Higgs boson, and the Higgsinos
to be at the gravitino mass scale.
The gaugino masses are assumed to be dominated by the anomaly mediation effects~\cite{Giudice:1998xp,Randall:1998uk} 
which are roughly given by (see, e.g.~\cite{Bhattacherjee:2012ed}),
\begin{eqnarray}
m_{\rm bino} &\simeq& 10^{-2}\times m_{3/2}\ , \\
m_{\rm wino} &\simeq& 3\times 10^{-3}\times m_{3/2}\ , \\
m_{\rm gluino} &\simeq& 2.5\times 10^{-2}\times m_{3/2}\ , 
\end{eqnarray}
although the constraints do not depend on them significantly as long as they are in the TeV range.
The gravitino mass is take to be within $10\,{\rm TeV}\le m_{3/2}\le 10$\,PeV.
These choices are motivated by the pure gravity mediation model in Refs.\,\cite{Ibe:2006de,*Ibe:2011aa,*Ibe:2012hu} 
(see also Refs.\,\cite{Giudice:2004tc,ArkaniHamed:2004yi,Wells:2004di,ArkaniHamed:2012gw} 
for closely related models).%
\footnote{Here, the Higgsino mediation effects neglected for simplicity.
Besides, the gaugino spectrum is deflected from the 
anomaly mediation in the presence of the KSVZ extra multiplets~\cite{Harigaya:2013asa}.}
In the renormalization group evolution, we also take into account an extra multiplet required for the anomaly free condition of the 
${\mathbb Z}_{4R}$ symmetry, whose masses are also at the gravitino mass scale.

The figure shows that the requirement for perturbative unification excludes the charges with $q_2> 7$ ($N_{f} > 7$).
This is expected as  $N_f$ flavors of the KSVZ extra multiplets have masses of $10\,{\rm TeV}\lesssim m_{KSVZ} \lesssim 10$\,PeV.%
\footnote{If we restrict to $m_{3/2}<1$\,PeV, the constraint becomes tighter and the charges with $q_2 >5$ are excluded.}
On the other hand, a large $q_1$ is allowed.
This is because the explicit breaking terms are suppressed by $(\Lambda'/M_{\rm PL})^{3q_1}$,
and hence, a high-quality global PQ is possible even  for a large $\Lambda'$ as long as $q_1$ is large.
For a large $\L'$,  $m_{KSVZ}'$ also becomes  large, with which the perturbative unification is  possible even if $N_f' = q_1$ is large.
It should be noted, however, that the effective field theory approach is no more reliable when $\Lambda'$ is too close to the Planck scale.
In the figure, we color the charges by orange if they require a large $\Lambda'$, i.e., $10^{16}\,{\rm GeV} \lesssim \Lambda' \lesssim 10^{17}$\,GeV.
For $N_{\rm GCD} \ge 2$, there are no appropriate charges 
with which  $\theta_{\rm eff}< 10^{-10}$ and the perturbative unification are compatible.

\subsection{Parameter Regions in the $SU(3)'$ Model}
\begin{figure}[t]
\begin{center}
\begin{minipage}{\linewidth}
  \includegraphics[width=\linewidth]{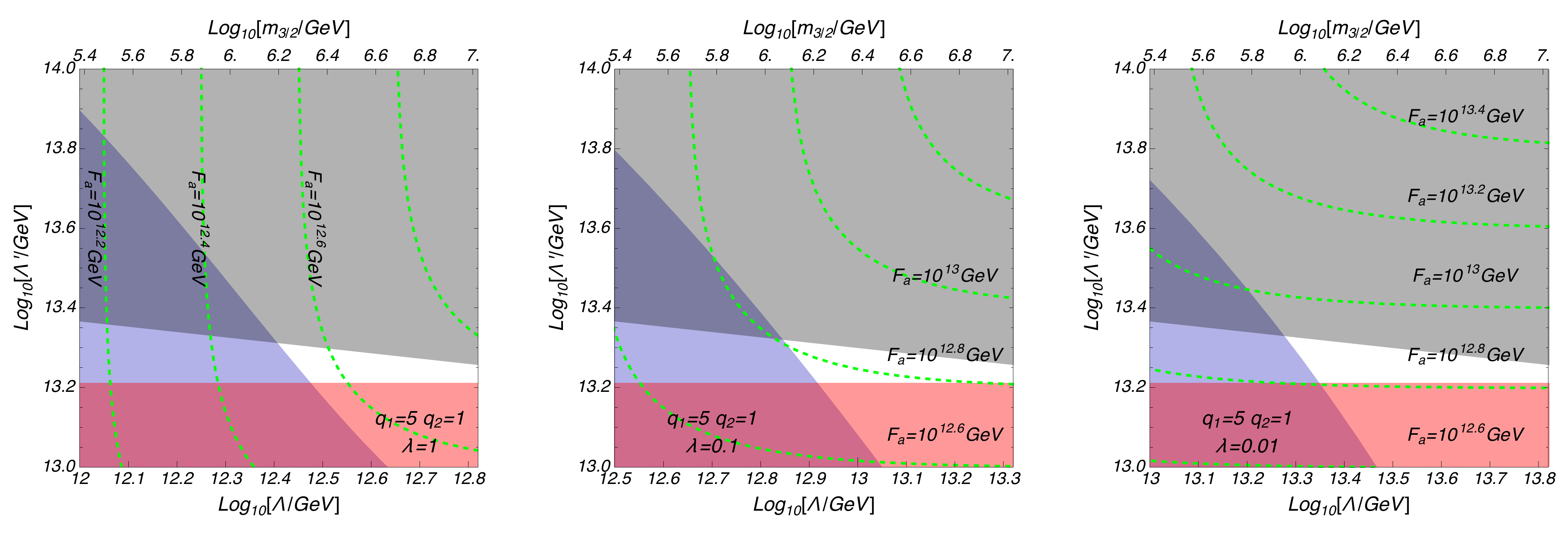}
 \end{minipage}
 \begin{minipage}{\linewidth}
  \includegraphics[width=\linewidth]{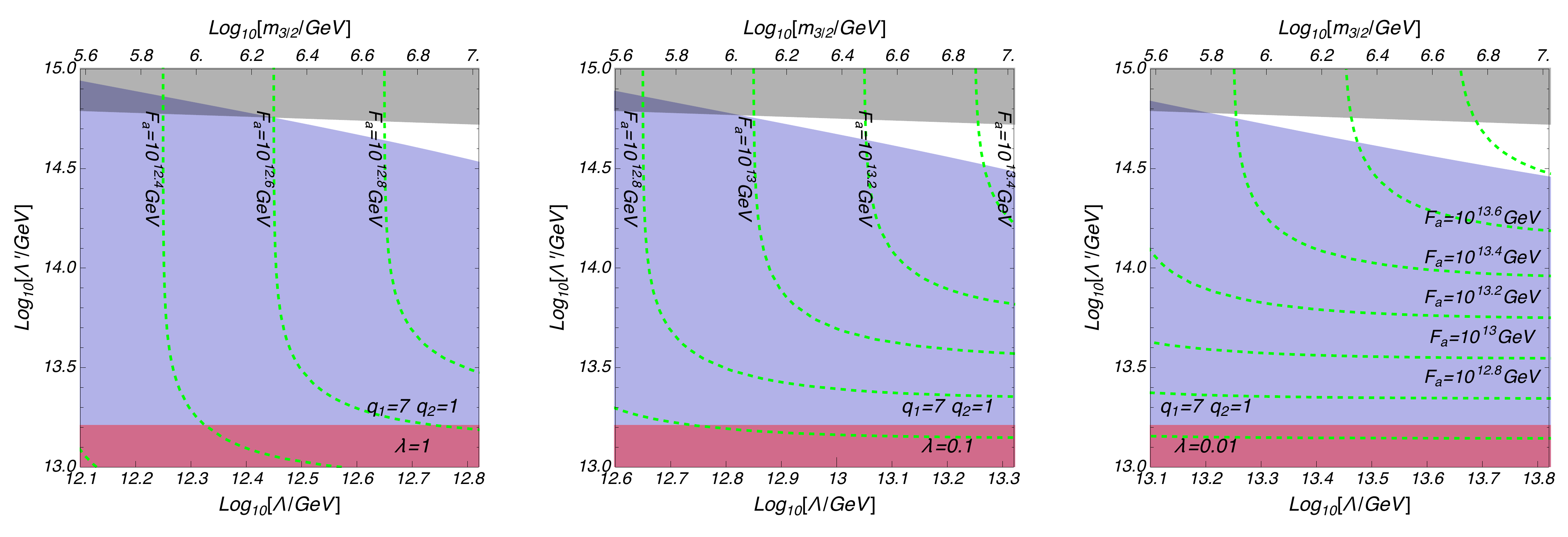}
 \end{minipage}
  \begin{minipage}{\linewidth}
  \includegraphics[width=\linewidth]{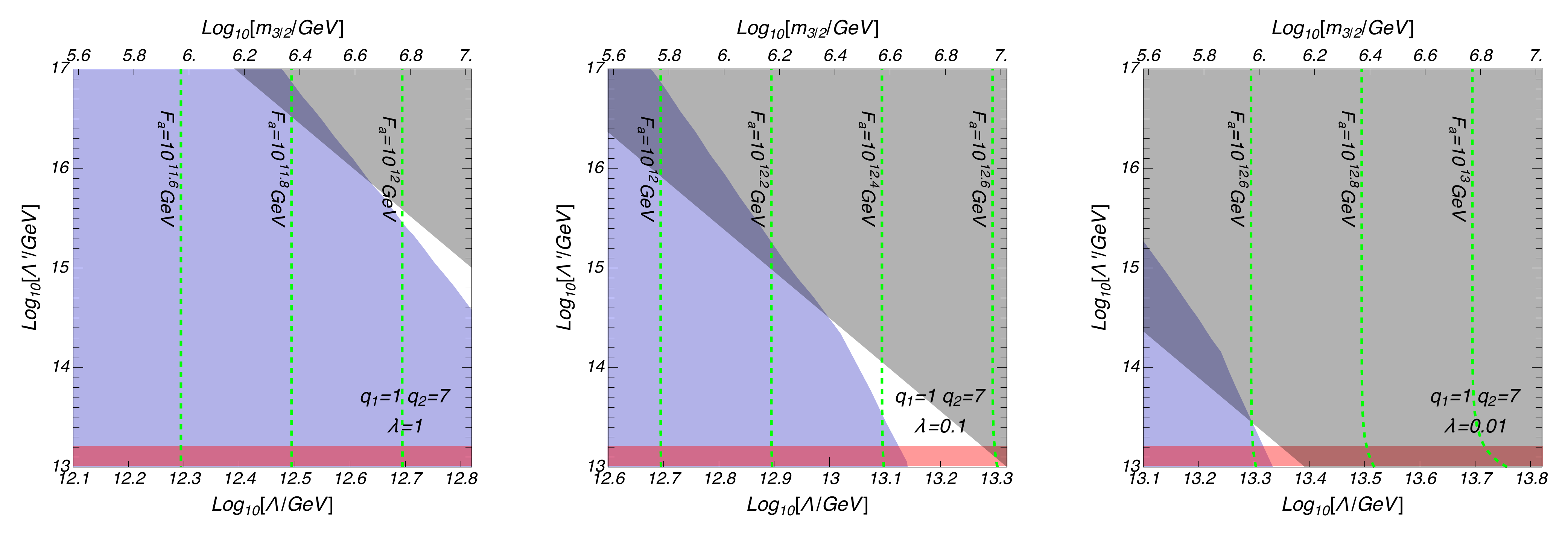}
 \end{minipage}
  \end{center}
\caption{\sl \small  
The constraints on the parameter reions for given PQ charges, $q_1$ and $q_2$.
The gray region is excluded as $\theta_{\rm eff} < 10^{-10}$ is not satisfied. 
The perturbative unification is not achieved in the blue region.
The red regions are excluded by $m_{KSVZ}' \gtrsim 750$\,GeV.
The green lines are the contours of the effective decay constant $F_a$.
}
\label{fig:parameters}
\end{figure}

In Fig.\,\ref{fig:parameters}, we  show the  parameter regions for a given $q_1$ and $q_2$.
In each panel, we take $m_{3/2}<10$\,PeV and $\lambda = 1$, $10^{-1}$, $10^{-2}$, respectively.
The gray shaded region is excluded, as $\theta_{\rm eff} < 10^{-10}$ is not satisfied (see Eq.\,(\ref{eq:theta3})).
The perturbative unification is not achieved in the blue shaded region.
The red shaded region is excluded for too light KSVZ extra multiplets, i.e., $m_{KSVZ}' \lesssim 750$\,GeV. 
The green dashed lines are contours of the effective decay constant in Eq.\,(\ref{eq:effF}).
 
The figure shows that  the dynamical scale $\Lambda'$ is tightly constrained from above to achieve $\theta_{\rm eff} < 10^{-10}$ 
for the minimum charge choice, i.e., $q_1=5$ and $q_2 =1$.
This is understood as the explicit breaking terms are not effectively suppressed for rather small charges.
As a result, the PQ breaking scales are required to be low to avoid large explicit breaking effects.
The upper limit on $\Lambda'$  becomes tighter for a larger $\Lambda$ as is expected from Eq.\,(\ref{eq:theta3}).
Furthermore, as the dynamical scale $\Lambda$ becomes larger for a smaller $\lambda$,  
the upper limit becomes even tighter for a smaller $\lambda$ for a given $m_{3/2}$.
The constraints from the perturbative unification are, on the contrary,  weaker since $m_{KSVZ}$ becomes larger for a smaller 
$\lambda$ for a given $m_{3/2}$.

An interesting property of the minimum choice is that the model predicts the KSVZ extra multiplets ($\mathbf 5'$, $\bar{\mathbf 5}'$)
in the TeV range.
This feature reflects the suppressed fermion masses of the KSVZ extra multiplet in Eq.\,(\ref{eq:KSVZp})
caused by the composite nature of the PQ breaking field, i.e., $B_-'$, with a tight upper limit on  $\Lambda'$.
Thus, the model with the minimum charge choice can be tested by searching for  vector-like colored particles at the LHC experiments.

For $q_1=7$ and $q_2 = 1$, the upper limit on $\Lambda'$ is weaker than for the minimum choice.
This is because the suppression factor of the explicit breaking term, $(\Lambda'/M_{\rm PL})^{3q_1}$,
can be very small even for a rather large $\Lambda'$ due to a large exponent.
The constraint form the perturbative unification is, on the contrary, tighter for a large $q_1$ as 
 $N_f'$ is proportional to $q_1$.
 For a large $N_f'$, the masses of the KSVZ extra multiplets, $m_{KSVZ}'$, is required to 
 be high to avoid the blow-up of the gauge coupling constants below the GUT scale.

For $q_1 =1$ and $q_2 = 7$, the upper limit on $\Lambda'$ is also weaker than the minimum choice for $\lambda = 1$
due to a strong suppression of the explicit breaking terms by $(\Lambda/M_{\rm PL})^{2 q_2}$.
As the suppression factor is sensitive to $\Lambda$, the upper limit on $\Lambda'$ becomes 
very tight for a smaller $\lambda$ for a given gravitino mass. 

In all cases, we find that the gravitino mass is required to be in the hundreds TeV or larger,
and hence, the model can be consistent with the observed Higgs boson mass
achieved by the gravity mediated sfermion masses.
It is also notable that  the dynamical scale $\Lambda'$ is larger than $\Lambda$ in the allowed parameter region.
Therefore, both the accidental global PQ symmetry and supersymmetry are broken by the IYIT sector
while the gauged PQ symmetry is mainly broken by the $SU(3)'$ sector.
This feature is attractive as it explains the coincidence between the global PQ breaking scale and the supersymmetry breaking scale.

Before closing this subsection, let us comment on the axion dark matter abundance.
The axion starts coherent oscillation when the Hubble expansion rate becomes comparable to the 
axion mass, which leads to the present axion dark matter density~\cite{Turner:1985si},
\beq
\Omega_{\rm axion}h^2\simeq 0.2\times \theta^2_{i}\left(\frac{F_a}{10^{12}\,{\rm GeV}}\right)^{1.19}\ .
\eeq
Here, $\theta_{i}$ is the initial misalignment angle of the axion field.
Thus, the axion can be a dominant component for dark matter of $F_a  = {\cal O}(10^{12})$\,GeV, i.e., $\Omega_{\rm DM} \simeq 0.12$~\cite{Calabrese:2017ypx}.
As the figures show,  $F_a = {\cal O}(10^{12})$\,GeV is possible in a wide range of the parameter space.
Therefore, the model based on $SU(3)'$ can be consistent with the axion dark matter scenario.%
\footnote{For $m_{3/2} \gg {\cal O}(1)$\,PeV, the wino is expected to be 
heavier than ${\cal O}(1)$\,TeV, whose relic abundance exceeds the observed dark matter density.
In such parameter region, we need to assume either a dilution mechanism of dark matter
or $R$-parity violation.
}

\subsection{Cancellation of  Self- and Gravitational Anomalies }
\label{sec:anomcancel}
As mentioned in section~\ref{sec:prescription}, 
the  gravitational anomaly and the self-anomaly of  $U(1)_{gPQ}$ are canceled 
by adding $U(1)_{gPQ}$ charged singlet fields.
In this subsection, we show a concrete model of the anomaly cancelation.

In the IYIT sector and the $SU(3)'$ sector, 
the $U(1)_{gPQ}$ charged fields are paired with fields with opposite charges.
Thus, the fields in these sectors do not contribute to the self-anomaly nor 
the gravitational anomaly.
The charges of the KSVZ extra multiplets are, on the other hand, not paired, and hence,
they contribute to the anomalies,
\begin{eqnarray}
{\cal A}_{\rm self}^{\rm KSVZ}&=&-5 N_f q_1^3   + 5 N_f'  q_2^3 \ , \\
{\cal A}_{\rm gravitational}^{\rm KSVZ} &=&-5 N_f q_1  +5 N_f' q_2  \ ,
\end{eqnarray}
respectively.
The easiest way to cancel the anomaly is to introduce $5N_f$ singlet superfields $Y$ 
with a charge $q_1$ and $5N_f'$ singlet superfields $Y'$ with
a charge $-q_2$.
The charges  of $Y$'s and $Y'$'s are given in Tab.~\ref{tab:gPQ}.

As the singlet fields do not have mass partners with opposite charges, the supersymmetric masses of them are 
generated only after $U(1)_{gPQ}$ breaking.
The mass terms of $Y$'s are given by
\beq
\label{eq:mY}
W\sim \frac{m_{3/2}}{M_{\rm PL}^4} (Q_3Q_4)^2YY \sim m_{3/2}\frac{\Lambda^4}{M_{\rm PL}^4}YY\ .
\eeq
Here, we take the ${\mathbb Z}_{4R}$ charge of $Y$'s to be $1$, so that their scalar and fermion components 
are odd and even under the $R$-parity, respectively. 
The factor  $m_{3/2}$ encapsulates the effects of  spontaneous breaking of the ${\mathbb Z}_{NR}$ symmetry.
As a result, the fermionic components of $Y$'s obtain
\begin{eqnarray}
m_Y \sim 3\,\times 10^{-7}\,{\rm eV} \left(\frac{m_{3/2}}{10^6\,{\rm GeV}}\right)\left(\frac{\Lambda}{10^{13}\,{\rm GeV}}\right)^4\ ,
\end{eqnarray}
while the masses of scalar components are dominated by the gravity mediated 
soft masses of ${\cal O}(m_{3/2})$.

The supersymmetric masses of  $Y'$'s are even smaller,
\beq
\label{eq:mYp}
W\sim \frac{m_{3/2}}{M_{\rm PL}^6} (\bar{Q}'\bar{Q}'\bar{Q}')^2Y'Y' \sim m_{3/2} \left(\frac{\Lambda'}{M_{\rm pl}}\right)^6Y'Y'\ .
\eeq
Here, we take the ${\mathbb Z}_{4R}$ charge of $Y'$'s to be $1$, and the 
factor $m_{3/2}$ encapsulates the effects of  spontaneous breaking of ${\mathbb Z}_{NR}$ again.
As a result, the fermionic components of $Y'$'s obtain,
\begin{eqnarray}
m_{Y'} \sim 5\times 10^{-12}\,{\rm eV}
\left(\frac{m_{3/2}}{10^6\,{\rm GeV}}\right)
\left(\frac{\Lambda'}{10^{14}\,{\rm GeV}}\right)^6
\end{eqnarray}
while the masses of the scalar components of $Y'$'s are  dominated by the gravity mediated soft masses
as in the case of $Y$'s.%
\footnote{There are mass terms  proportional to $YY'$ which can be lager than Eqs.\,(\ref{eq:mY}) and (\ref{eq:mYp})
depending on the parameters.
Even in such cases, there remain light fermions with masses either $m_Y$ or $m_Y'$  as the numbers of $Y$'s and $Y'$'s are different.
}

If the light fermions are abundantly  produced in the early universe, they contribute to the dark radiation and result in
an unacceptably large number of effective neutrino species, $N_{\rm eff}$. 
To evade this problem, we assume that spontaneous breaking of $U(1)_{gPQ}$ takes place before the end of inflation.
We also assume that the gauge superfields of $U(1)_{gPQ}$ are heavier than the reheating temperature after inflation.
Furthermore, it is also assumed that the branching fraction of the inflaton into $Y$'s and $Y'$'s are suppressed.
With these assumptions, we can achieve cosmologically consistent models where the self- and the gravitational anomalies 
are canceled by the $U(1)_{gPQ}$ charged singlets.

\subsection{$SU(N)'$ Dynamical PQ Symmetry Breaking Model}
So far, we have considered the dynamical PQ breaking sector based on the $SU(3)$ gauge theory.
There, the deformed moduli constraint plays an important role to break the global PQ symmetry (i.e., the baryon symmetry)
spontaneously.
In this subsection, we discuss the models of dynamical PQ breaking based on $SU(N)$  gauge theory other than
$N = 3$.
We call such models, the $SU(N)'$ dynamical PQ breaking model.

First, let us consider the $SU(2)'$ model.
With four fundamental representations of $SU(2)'$, $Q'$, the model exhibits the deformed moduli constraint.
In this model, there is no baryon symmetry, and the global PQ symmetry is identified with a subgroup 
of the maximal non-abelian group $SU(4)_f$ as in the case of the IYIT sector.
Then, the global PQ symmetry breaking is achieved 
by introducing four PQ neutral singlet superfields, $Z'$.%
\footnote{It is tempting to make the $SU(2)'$ sector also be the IYIT supersymmetry breaking sector by introducing six singlet fields, $Z'$'s, instead.
In this case, however, supersymmetry and the gauged PQ symmetry are broken by the dynamics, while the global PQ symmetry is broken separately.}

In this model, the KSVZ extra multiplets coupling to the $SU(2)'$ sector obtain masses via,
\begin{eqnarray}
W\sim \frac{1}{M_{\rm PL}} Q_1' Q_2'   {\mathbf 5}'\,\bar{\mathbf 5}' \ ,
\end{eqnarray}
leading to
\begin{eqnarray}
m_{KSVZ'} \sim \frac{\Lambda'^2}{M_{\rm PL}}\ .
\end{eqnarray}
Thus, the $SU(2)'$ model allows a rather small $\Lambda'$ compared with the $SU(3)'$ model to achieve $m_{KSVZ}' \gtrsim 750$\,GeV.
It is even possible to be $\L' \ll\L$.
The possibility of $\L' \ll \L$ is, however, not very attractive as the model does not 
explain  a  coincidence between the PQ breaking scale and the supersymmetry breaking scale.

Next, let us consider the $SU(N)'$  $(N>3)$ model.
In this case, the global PQ symmetry is identified with the baryon symmetry which is broken by
the deformed moduli constraint as in the case of the $SU(3)'$ model.
As the mass terms of the KSVZ extra multiplets, $\mathbf 5'$ and $\bar{\mathbf 5}'$, are given by,
\begin{eqnarray}
W\sim \frac{1}{M_{\rm PL}^{N-2}} (\bar{Q}'\cdots \bar{Q}')   {\mathbf 5}'\,\bar{\mathbf 5}' \ ,
\end{eqnarray}
the dynamical scale $\Lambda'$ should be much higher than $\Lambda$ to satisfy $m_{KSVZ}' \gtrsim 750$\,GeV.
Here, $\bar{Q}\cdots \bar{Q}$ denotes the baryon operators of the $SU(N)'$ sector.
The $SU(N)'$ models are very similar to the $SU(3)'$ model except for the dynamical scale $\Lambda'$, 
although we do not discuss details of the $SU(N)'$ model further.

\section{Conclusions}
\label{sec:cons}
In this paper, we apply the gauged PQ mechanism to a model in which the global PQ symmetry and supersymmetry are broken simultaneously.
As a concrete example, we considered models which consist of simultaneous supersymmetry/PQ symmetry breaking sector based on $SU(2)$ dynamics
(the IYIT sector) and a dynamical PQ symmetry breaking sector based on $SU(N)$ dynamics (the $SU(N)'$ sector).
As we have seen, the $SU(3)'$ model is particularly successful where the gauged PQ symmetry is mainly broken by the $SU(3)'$ sector
while both the accidental global PQ symmetry and supersymmetry are broken by the IYIT sector.
Thus, the model explains the coincidence between the favored values of the supersymmetry breaking scale and the PQ breaking scale.

Besides the model with the minimum charge choice, $q_1 = 5$ and $q_2 = 1$, predicts the KSVZ extra multiplets ($\mathbf 5'$, $\bar{\mathbf 5}'$)
in the TeV range due to a tight upper limit on $\Lambda'$.
It should be noted that the light KSVZ fermions are due to composite nature of the PQ breaking field
with a tight upper limit on $\L'$ in the minimal model.
The model also predicts non-vanishing effective $\theta$ angle. 
Thus, the model with the minimum charge choice can be tested by combining the searches for vector-like colored particles at the LHC experiments
and future measurements of the neutron EDM.

Finally, let us comment an advantage of the gauged PQ mechanism over the models in which 
the high-quality global PQ symmetry results from an exact discrete symmetry, such as ${\mathbb Z}_N$.
As we have discussed briefly in subsection~\ref{sec:domainwall}, the gauged PQ mechanism with $N_{\rm GCD} = 1$,  
$q_1 = 1$ and $q_2 = N (>1)$ allow models which are free from both the domain wall problem and the axion isocurvature problem. 
The assumption here is that the first stage of the phase transition (i.e. $\vev{\Phi_1}\neq 0$) takes place 
before inflation while the second stage of the phase transition (i.e. $\vev{\Phi_2} \neq 0$) occurs after inflation.
Then, the local strings formed at the first phase transition are inflated away,
while the global strings formed at the second phase transition do not cause 
the domain wall problem as $\Phi_2$ couples to only one-flavor of the KSVZ extra multiplet.%
\footnote{In this case, the axion dark matter density is dominated by the axions produced
by the decay of the string-domain wall networks, which requires $F_a = {\cal O}(10^{11})$\,GeV~\cite{Hiramatsu:2010yn,Hiramatsu:2012gg}. 
Such a rather low $F_a$ is, for example, achieved in the $SU(2)'$ model.
}
As the global PQ symmetry is broken after inflation, the model does not suffer from the axion isocurvature problem.
This option is not available in the models with an exact discrete symmetry where the axion potential is also symmetric
under the discrete symmetry.

\begin{acknowledgments}
The authors would like to thank K.~Harigaya and K.~Yonekura for useful discussion on domain wall problems.
This work is supported in part by Grants-in-Aid for Scientific Research from the Ministry of Education, Culture, Sports, Science, and Technology (MEXT) KAKENHI, Japan, No. 25105011,  No. 15H05889 and No. 16H03991 (M. I.),  No. 26104001 and No.26104009 and No.16H02176  (T. T. Y.); 17H02878 (M. I. and T. T. Y.), 
and by the World Premier International Research Center Initiative (WPI), MEXT, Japan.
The work of H.F. is supported in part by a Research Fellowship for Young Scientists from the Japan Society for the Promotion of Science (JSPS).
\end{acknowledgments}

\bibliography{papers}

\end{document}